\documentclass[%
 reprint,
superscriptaddress,
 amsmath,amssymb,
 aps,
 prl
]{revtex4-2}

\usepackage{graphicx}
\usepackage{dcolumn}
\usepackage{bm,upgreek}
\usepackage{soul}
\usepackage[normalem]{ulem}

\usepackage{color}
\usepackage{physics}
\usepackage{hyperref}

\usepackage{bibunits}

\begin{document}

\newcommand{\papertitle}{Nonlinear effects in Anderson localization of light by two-level atoms}
\title{ \papertitle}

\author{Noel Araujo Moreira}
\affiliation{Instituto de F\'isica de S\~{a}o Carlos, Universidade de S\~{a}o Paulo - 13566-590 S\~{a}o Carlos, SP, Brazil}

\author{Robin Kaiser}
\affiliation{Universit\'e C\^ote d'Azur, CNRS, INPHYNI, France}

\author{Romain Bachelard}
\email{romain@ufscar.br}
\affiliation{Universit\'e C\^ote d'Azur, CNRS, INPHYNI, France}
\affiliation{Departamento de F\'{\i}sica, Universidade Federal de S\~{a}o Carlos, Rodovia Washington Lu\'{\i}s, km 235 - SP-310, 13565-905 S\~{a}o Carlos, SP, Brazil}

\date{\today}

\begin{abstract}
While Anderson is a single-particle wave effect, guaranteeing a single excitation in the system can be challenging. We here tackle this limitation in the context of light localization in three dimensions in disordered cold atom clouds, in presence of several photons. We show that the presence of these multiple excitations does not affect substantially the abnormal intensity fluctuations which characterize the Anderson localization transition, provided that the radiated light is frequency filtered. Due to their narrow linewidth, long-lived modes, and particularly the localized ones, are strongly saturated even for a weak resonant pump, leading to a large increase of the inelastic scattering and to reduced fluctuations in the total radiation. Yet the atomic coherences and the resulting elastic scattering remain a proper witness of the Anderson localization transition. Hence, frequency filtering allows one to investigate the single-excitation sector, dismissing the many-body effects showing up in the fluorescence spectrum.
\end{abstract}

\maketitle


{\em Introduction.---} Since its introduction in the context of the metal-to-insulator transition for electronic transport~\cite{Anderson1958}, Anderson localization by disorder has been shown to be a general wave phenomenon. In three dimensions, it has since been reported experimentally for elastic waves~\cite{Hu2008}, atomic matter waves~\cite{Chabe2008}, and electrons~\cite{Ying2016}. Anderson localization formally corresponds to the localization of a single excitation, that is, if several waves/excitations are present, they do not interact with each other. 

In this context, the localization of light seems particularly promising, since photons are notoriously inefficient at interacting with each other. However, the initial experimental reports of light localization in 3D~\cite{Wiersma1997,Storzer2006} have been later reinterpreted~\cite{Scheffold1999,Scheffold2013,Sperling2016}, and an unambiguous observation is still missing~\cite{Skipetrov2016}. From a theoretical point of view, the near-field terms coupling the different polarization channels have been pointed at as an obstacle to localization~\cite{Skipetrov2014,Naraghi2016,Cobus2022}, challenging the mere existence of Anderson localization of light in three dimensions. These advances stimulated new proposals to restore light localization~\cite{Skipetrov2015,Celardo2017}, taking advantage of the tunability of the light-emitter interaction in cold atom platforms, along with their relative absence of decoherence mechanisms: Decoupling the polarization channels with a strong external field~\cite{Skipetrov2015}, or randomly shifting the atomic resonance of each atom with a disordered field~\cite{Celardo2017} -- one step closer from the original Anderson model~\cite{Anderson1958}. 

Nonetheless, while the use of a weak drive may seem sufficient to guarantee that the atoms will react linearly to the pump field, thus guaranteeing the pristine condition of non-interacting waves, long-lived modes have actually been reported to be particularly sensitive to nonlinear effects~\cite{Williamson2020,Cipris2021}. Indeed, localized modes present lifetimes which are order of magnitude larger than that of single atoms, so their effective saturation may be equally larger. This questions the possibility to observe light localization in cold atomic clouds, as nonlinear effects may arise even for the weakest pumps.

In this work, we investigate the localization of light in disordered clouds of two-level atoms submitted to a classical weak pump. A mean-field (MF) approximation allows us to simulate large disordered systems, neglecting the quantum correlations between the atomic dipoles, yet capturing collective linewidths and frequency shifts. Close to the atomic resonance, where localized and subradiant modes are most efficiently addressed, a strong increase of the inelastic scattering is observed, which stems from the narrow linewidth of these modes. Nevertheless, we show that the coherence stored in the atomic dipoles preserves the signature of the localization transition, even when the localized modes are strongly saturated: The resulting enhanced intensity fluctuations can be monitored by frequency filtering the radiated light, as the elastically-scattered signal exhibits these abnormal statistics~\cite{Chabanov2000,Cottier2019}. Our work is thus a first step toward the transition from single- to multi-excitation localization of light in three-dimensional cold atom systems.

{\em Single-Excitation vs. Mean-Field.---} Let us here consider a cloud of $N$ two-level atoms (ground and excited states $\ket{g_j}$ and $\ket{e_j}$, respectively) with positions $\mathbf{r}_j$, with a transition characterized by its frequency $\omega_a$, linewidth $\Gamma$, and raising/lowering operators $\hat\sigma_j^\pm$ for atom $j$. The system is pumped by a monochromatic classical field with a Gaussian profile of waist $w_0$, Rabi frequency $\Omega_0$ at the waist, and detuned by $\Delta=\omega_\textrm{laser}-\omega_a$ from the atomic transition. We introduce the resonant saturation parameter at the beam waist: $s_0=2\Omega_0^2/\Gamma^2$, and the non-resonant one $s(\Delta)=2\Omega_0^2/(\Gamma^2+4\Delta^2)$. Within the Born-Markov approximation, the light-mediated interactions between the atomic dipoles gives rise to a master equation of the form $ d\rho/dt = -\frac{i}{\hbar}\mathcal{H}[\rho] + \mathcal{L}[\rho]$
associated to the following Hamiltonian and Lindbladian:
\begin{eqnarray}
\mathcal{H}[\rho] = -\frac{i}{\hbar}\sum_{j=1}^N[H_j, \rho] - i\sum_{j, m \ne j}^N\Delta_{jm}[\hat\sigma^+_j\hat\sigma^-_m, \rho],\\
\mathcal{L}[\rho] = \frac{1}{2}\sum_{j,m} \Gamma_{jm}[2\hat\sigma^-_j\rho \hat\sigma^+_m - \hat\rho\sigma^+_m\hat\sigma^-_j - \hat\sigma^+_m\hat\sigma^-_j \rho],
\end{eqnarray}
with $H_j = -\frac{\hbar}{2} \Delta_j \hat\sigma^z_j + \frac{\hbar}{2}\Omega_j(\hat\sigma_j^+ + \hat\sigma_j^-)$ the single-atom Hamiltonian term ($\Omega_j$ the local Rabi frequency), $\Delta_{jm}=-(\Gamma/2) \cos(k_0|\mathbf{r}_j-\mathbf{r}_m|)/k_0|\mathbf{r}_j-\mathbf{r}_m|$ and $\Gamma_{jm}=\delta_{jm}\Gamma+(1-\delta_{jm})\Gamma \sin(k_0|\mathbf{r}_j-\mathbf{r}_m|)/k_0|\mathbf{r}_j-\mathbf{r}_m|$ the dipole-dipole interaction terms~\cite{Lehmberg1970}. Here we work in the scalar wave approximation, in which localization occurs without resorting to external fields~\cite{Skipetrov2015,Celardo2017}.

Anderson localization refers formally to single-excitation dynamics, when the waves do not interact with each other. This corresponds to the case of states with at most one photon, of the form $\ket{\psi}=\alpha\ket{g_1g_2..g_N}+\sum_{j=1}^N\beta_j\ket{g_1..e_j..g_N}$, which leads to the following equation describing the evolution of the atomic coherences:
\begin{equation}
    \frac{d\beta_j}{dt} = \left ( i\Delta  -\frac{\Gamma}{2} \right ) \beta_j - i\frac{\Omega_j}{2} -\frac{\Gamma}{2} \sum_{m \ne j}^N  \frac{e^{ik_0|\mathbf{r}_j-\mathbf{r}_m|}}{ik_0|\mathbf{r}_j-\mathbf{r}_m|} \beta_m,\label{eq:CDE}
\end{equation}
hereafter referred to as Coupled Dipole Equations (CDE). The set of equations~\eqref{eq:CDE} describes the optical coherences, $\beta_j=\langle\hat\sigma_j^-\rangle$, and contains no information on the excited population. In order to investigate the role of a finite pump strength, and thus the presence of multiple photons in the system, we resort to the MF approach. It accounts for the finite atomic population, $z_j=\langle\hat\sigma_j^z\rangle$, and neglects connected correlations between the atoms: $\langle\sigma_j^\alpha \sigma_m^\beta\rangle\equiv \langle\sigma_j^\alpha\rangle \langle \sigma_m^\beta\rangle$. We then obtain the following $2N$ equations for the coherences $\beta_j$ and the populations $z_j$~\cite{Santo2020}:
\begin{subequations}\label{eq:MF}
\begin{align} 
    \frac{d \beta_j}{dt} &=\left (i\Delta - \frac{\Gamma}{2} \right ) \beta_j + iW_jz_j,\label{eq:mf_sigma_minus} \\
   \frac{d z_j}{dt}  &= -\Gamma(1 + z_j) - 4\Im(\beta_j W_j^*),\label{eq:mf_sigma_z}
\\       W_j &= \frac{\Omega_j}{2} -\frac{\Gamma}{2}\sum_{m \ne j}^N \frac{ e^{ik_0|\mathbf{r}_j-\mathbf{r}_m|}}{ik_0|\mathbf{r}_j-\mathbf{r}_m|}\beta_m,
\end{align}
\end{subequations}
where $W_j$ is the effective Rabi frequency for atom $j$, composed of the pump and of the radiation from other atoms.
The MF approximation is necessary to reduce drastically the complexity of the Hilbert space of a 3D system, yet accounting for the saturation of the atoms. In particular, the set of equations~\eqref{eq:MF} is nonlinear, as one enters the realm of nonlinear optics where waves can interact with each other through the atomic medium.

The far-field intensity of the light scattered by the atoms in a direction $\hat{n}$, $I_\mathrm{tot}=\sum_{j,m}e^{-ik\hat{n}.(\mathbf{r}_j-\mathbf{r}_m)}\langle\hat\sigma_m^+\hat\sigma_j^-\rangle$, can be decomposed into an elastically- and inelastically-scattered components, $I_\mathrm{tot}=I_\mathrm{el}+I_\mathrm{in}$, given by:
\begin{subequations}
\begin{align} 
 I_\mathrm{el}&=\bigg|\sum_{j}e^{-ik\hat{n}.\mathbf{r}_j} \beta_j\bigg|^2,\label{eq:Iel}
 \\ I_\mathrm{in}&=\sum_j \frac{1+z_j}{2}-\left|\beta_j\right|^2 .\label{eq:Iin}
\end{align} 
\end{subequations}
Note that the inelastic component~\eqref{eq:Iin} contains only single-atom contributions due to the MF approximation, which neglects two-atom connected correlations.

\begin{figure}[t!]
	\centering	\includegraphics[width=1\columnwidth]{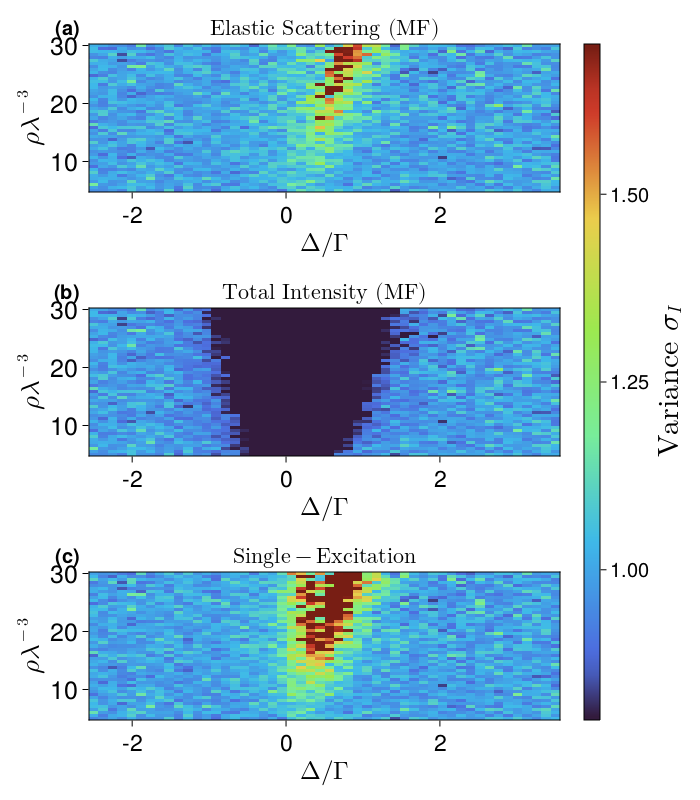}
	\caption{Variance of the scattered intensity, computed from its fluctuations over the azimuthal angle and over 50 realizations (a) for the elastically-scattered light $I_\textrm{el}$, (b) for the total intensity $I_\textrm{el}+I_\textrm{in}$, and (c) in the single-excitation regime~\eqref{eq:CDE}. Simulations realized for a cylindrical cloud with radius $R = 3\lambda$ and length $h = 6\lambda$, pumped by a Gaussian beam with waist $w_0 = 1.5\lambda$ and saturation parameter  $s_0=0.1$.}
	\label{fig:mapfluct}
\end{figure}

{\em Intensity fluctuations from saturated atoms.---} Intensity fluctuations have been reported to witness the Anderson localization transition in the single excitation regime~\cite{Chabanov2000,Cottier2019}: Let us now probe these in the multiple-excitation regime, using the MF approach~\eqref{eq:MF}. In Fig.~\ref{fig:mapfluct}, the evolution of the intensity variance of the scattered light, $\sigma_I=\langle I^2\rangle/\langle I\rangle^2$, computed over different azimuthal angles and realizations, is presented for a saturation parameter $s_0=0.1$: Panel (a) depicts the fluctuations of the elastically-scattered intensity $I_\textrm{el}$ in a range of detuning and density for which localization manifests [see panel (c) and Ref.~\cite{Cottier2019} for the single-excitation case]. The radiation from the coherences~\eqref{eq:Iel} thus presents large intensity fluctuations in the localization region, despite the presence of multiple excitations in the system. Indeed, at first order the number of excitations $N_e$ in the cloud can be evaluated making an independent scatterer hypothesis where the cloud holds $N_e=Ns/2(1+s)$ excitations. In the case of Fig.~\ref{fig:mapfluct}, this corresponds to more than a hundred excitations. 

The total intensity scattered $I_\textrm{el}+I_\textrm{in}$ presents a very different behaviour, with reduced fluctuations close to resonance, see panel (b). This feature stems from the nature of the fluctuations investigated here: As mentioned before, these fluctuations refer to variations over space (azimuthal angle) and  atomic realizations of the intensity computed as an expected value $I\sim\langle\hat{E}^\dagger \hat{E}\rangle$. For a given direction of observation and a given realization, this expected value formally corresponds to an infinitely-long measurement for static atoms. Practically, this measurement needs to be long enough to capture a large number of photons and get a statistically representative intensity, yet short enough to prevent the loss of coherence from mechanisms such as the atomic motion~\cite{Lassegues2023}. However, spontaneous emission from the excited state brings in a new time scale, that is, the excited state lifetime $1/\Gamma$. A proper detection of the  fluctuations in the spontaneously emitted field require a treatment which addresses quantum fluctuations, so phenomena such as photon bunching and antibunching on a time scale $1/\Gamma$ can be addressed. In the context of the master equation used here, this would mean dealing with higher-order atom-atom correlations~\cite{Pucci2017,Plankensteiner2022}, which is beyond the scope of our work.

\begin{figure}[t!]
	\centering
	\includegraphics[width=1\columnwidth]{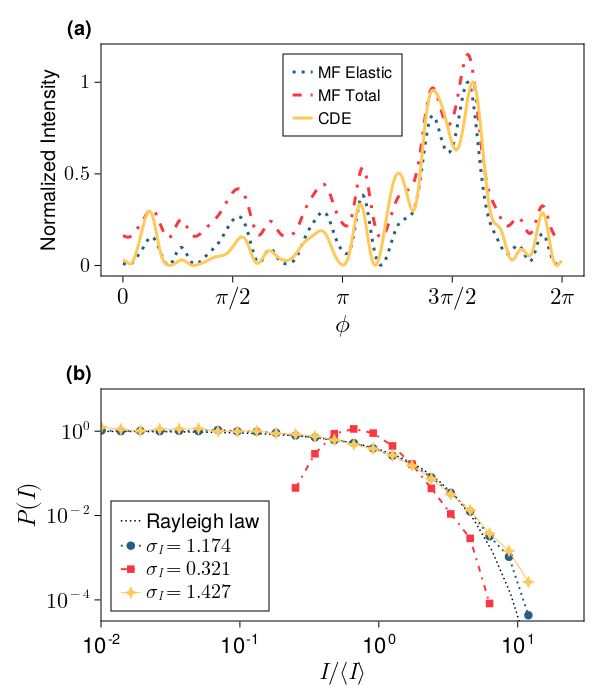}
	\caption{(a) Intensity as a function of the azimuthal angle $\phi$, considering the elastic component, the total field, and the single-excitation sector. (b) Intensity probability distribution function $P(I)$ for the same parameters, averaged over 100 realizations. Simulations realized for a spherical cloud  of $N=1500$ particles, density $\rho  = 25/\lambda^3$, detuning $\Delta = 0.8\Gamma$ and saturation parameter $s_0=0.1$.}
	\label{fig:stats}
\end{figure}

Thus, the measurement we consider does not capture fluctuations stemming from spontaneous emission, and the inelastic scattering contributes an homogeneous background for the radiated light. This is illustrated in Fig.~\ref{fig:stats}(a) where the polar profile of the intensity is plotted. The total intensity (red dash-dotted curve) presents the same fluctuations as the elastic component, yet shifted by the inelastic homogeneous background. 
Note that the linear CDE presents slightly different fluctuations from the MF approach: This is a signature of the saturation of the atoms, that is, of the excited population, which is accounted for in that model. The associated intensity probability density functions are represented in panel (b), where both the elastic component of the MF and the single-excitation signal exhibit increased fluctuations, with tails larger than for Rayleigh law, $P(I)\sim\exp(-I/\langle I\rangle)$, valid for uncorrelated scatterers. We note that these tails are responsible for the enhanced fluctuations in presence of Anderson localization~\cite{Cottier2019}. Due to the inelastic background, the total intensity explores, relatively, a smaller range of values, which results in reduced fluctuations. 

\begin{figure}[t!]
	\centering
 \includegraphics[width=1\columnwidth]{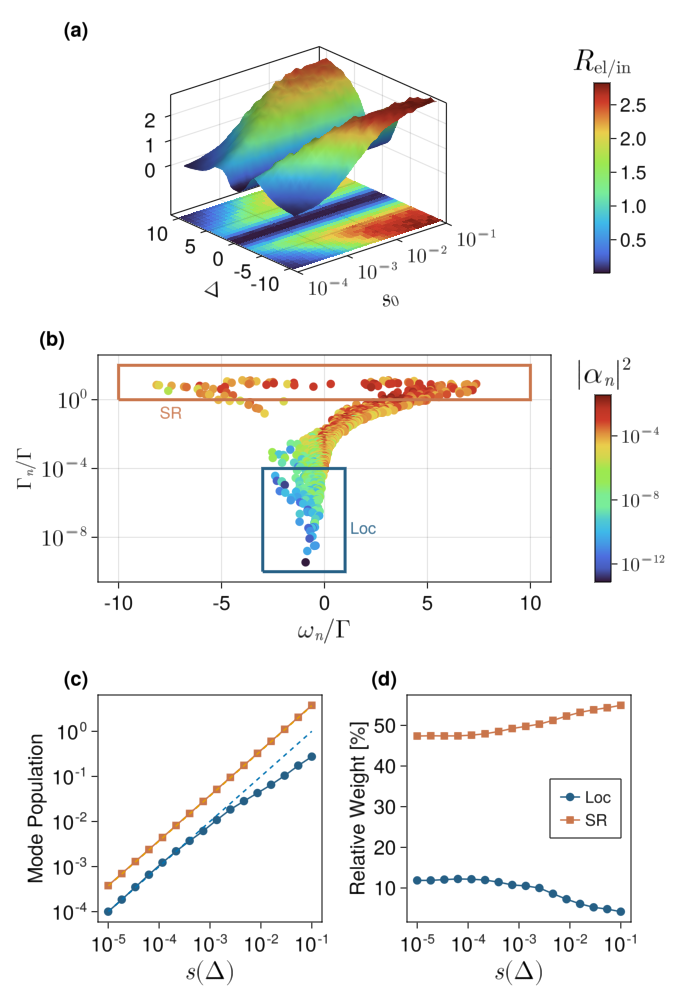}
	\caption{(a) Normalized ratio of elastically to inelastically scattered power $R_\textrm{El/In}=s(\Delta)P_\textrm{el}/P_\textrm{in}$, as a function of $\Delta$ and $s_0$, for a density $\rho=25/\lambda^3$. (b) Population of each mode, $|\alpha_n|^2$, in the complex plane of eigenvalues $\lambda_n=i\omega_n-\Gamma_n$, for a drive with $s_0=10^{-1}$ and $\Delta=0.5\Gamma$. The upper and rectangle encompass the localized (Loc) and superradiant (SR) states, respectively.  (c) Evolution of the population of localized and superradiant states in the atomic coherences, as a function of the saturation parameter $s$. Simulation realized for a spherical cloud with $N=1000$ particles, density $\rho = 146/\lambda^3$, and a drive detuned by $\Delta = 1.27\Gamma$, which corresponds to the typical energy of localized modes for these parameters), averaged of 20 realizations. (d) Relative weight of the localized modes, defined as $(\sum_{n\in\textrm{Loc/SR}}|\alpha_n|^2)/\sum_{n\in\textrm{All}}|\alpha_n|^2$, as a function of the saturation parameter $s(\Delta)$. Same parameters as for (c).}.
	\label{fig:satmodes}
\end{figure}

The fact that the localized modes are able to contribute substantially with elastically scattered light, and that strong intensity fluctuations can be observed, is not trivial: Indeed, long-lived modes are saturated even for low saturation parameters due to their narrow linewidth~\cite{Williamson2020,Cipris2021}. 
This is confirmed by the spectral analysis of the scattered power, monitoring the elastically and inelastically scattered powers:
\begin{eqnarray}
P_\mathrm{el} =  4\pi \sum_{j, m} \frac{\sin(k_0|\mathbf{r}_j-\mathbf{r}_m|)}{k_0|\mathbf{r}_j-\mathbf{r}_m|}\beta_j \beta_m^*, \\ 
P_\mathrm{in} = 4\pi\left[\sum_{j} \frac{1+z_j }{2} - |\beta_j|^2 \right].
\end{eqnarray}
These are obtained by integrating the intensity over all angles, and for independent scatterers the ratio between them is simply given by the saturation parameter: $P_\textrm{el}/P_\textrm{in}=1/s(\Delta)$. We thus define the ratio $R_\textrm{el/in}=s(\Delta)P_\textrm{el}/P_\textrm{in}$, which quantifies the inelastic contribution beyond the single atom effect. Its behavior as a function of the detuning and saturation parameter is presented in Fig.~\ref{fig:satmodes}(a): Close to resonance, where most long-lived modes are encountered and populated~\cite{Guerin2017}, spontaneous emission is stronger than for independent scatterers, which can be interpreted as the fact that even relatively low saturation parameters ($s\sim 10^{-4}$) are able to saturate the localized modes and make them radiate inelastically. Oppositely, far from resonance, the broad-linewidth superradiant modes are less saturated than independent scatterers would be, which in turn results in a stronger elastic scattering, yielding a ratio $R_\textrm{el/in}>1$.

Delving farther into collective scattering modes, we investigate the contribution of the localized modes to the optical coherences $\beta_j$. The modes are considered to be localized when their spatial shape presents an exponential decaying profile (more precisely, when the logarithm of their profile presents a linear decay with a $R^2$ Pearson parameter larger than $0.5$~\cite{Moreira2019}). We then decompose the vector of the steady-state coherences $\beta_j$ onto the basis of eigenvectors from the single-excitation sector, (that is, the eigenvectors $\hat{\Psi}_n$ of the scattering matrix of \eqref{eq:CDE}) as $\sum_n\alpha_n\hat{\Psi}_n$, and define the weight of the each mode in the coherence vector as $|\alpha_n|^2$. The map of this population is presented in Fig.~\ref{fig:satmodes}(b), in the complex plane of eigenvalues $\lambda_n=i\omega_n-\Gamma_n$, with $\gamma_n$ the mode inverse lifetime and $\omega_n$ its shift from the atomic resonance -- superradiant modes thus correspond to $\gamma_n>\Gamma$. Localized modes are weakly populated compare to superradiant ones, yet their large number makes up for their weak coupling to the external world.

Let us now define the weight of the localized (superradiant) modes in the coherences as $W_\textrm{Loc}=\sum_{n\in \textrm{Loc}}|\alpha_n|^2$ ($W_\textrm{SR}=\sum_{n\in \textrm{SR}}|\alpha_n|^2$). As shown in Fig.~\ref{fig:satmodes}(c), a below-linear growth of the weight with the saturation parameter $s_0$ is observed for localized modes.
This is yet another evidence that the localized modes are more easily saturated than superradiant ones, so their population grow slower with the saturation parameter. Note that this situation is different from the decay dynamics probed in Ref.~\cite{Cipris2021}, where the decay by collective spontaneous emission from multiple excitation states toward few excitation ones actually increases the contribution of long-lived states to the radiative dynamics: While the difference between spontaneously-emitted light and the coherently-scattered one was not done in that work, we here focus on the coherences, in the steady-state regime. In particular, monitoring the \textit{relative weight} of localized/superradiant modes in the coherence vector (by normalizing the vector of $\{\alpha_n\}$), the relative contribution of the localized ones is reduced by a factor three as the saturation parameter increases by four orders of magnitude, for the benefit of superradiant ones, see Fig.~\ref{fig:satmodes}(d) -- the remaining population lies in subradiant extended modes~\cite{Moreira2019}. The present observation of largely saturated localized modes even at low saturation parameter makes it all the more remarkable that the elastic-scattering intensity fluctuations characteristic of the localization transition be preserved for a finite-strength drive.

{\em Perspectives.---} Although localized modes are effectively strongly saturated by relatively weak probes, for which single atoms would remain very close to the ground state, the signature of intensity fluctuations at the localization transition is preserved, provided that the light is filtered to select the elastic scattering component. This result is particularly important for setups where the single-photon condition -- the pristine condition for Anderson localization of light -- is challenging to achieve.

Our work paves the way to future studies on light scattering in presence of multiple excitations~\cite{Rubies2023}. In particular, the fluorescence spectrum of collective modes may also reveal precious information regarding the correlations between the dipoles~\cite{Pucci2017}. Hence, delving deeper in the hierarchy of quantum correlations is a next natural step to understand the many-body regime of these disordered systems~\cite{Fayard2021}.

\begin{acknowledgments}
R.B. acknowledges the financial support of the São Paulo Research Foundation (FAPESP) (Grants No. 2018/15554-5, 2019/13143-0 and 2022/00209-6), from the Brazilian CNPq (Conselho Nacional de Desenvolvimento Científico e Tecnológico), Grant No. 313886/2020-2, and from the French government, through the UCA J.E.D.I. Investments in the Future project managed by the National Research Agency (ANR) with the reference number ANR-15-IDEX-01. RB and RK received support from the project STIC-AmSud (Ph879-17/CAPES 88887.521971/2020-00), CAPES-COFECUB (CAPES 88887.711967/2022-00) and from the European project ANDLICA, ERC Advanced grant No. 832219.
\end{acknowledgments}

\bibliography{Biblio}

\begin{thebibliography}{28}%
\makeatletter
\providecommand \@ifxundefined [1]{%
 \@ifx{#1\undefined}
}%
\providecommand \@ifnum [1]{%
 \ifnum #1\expandafter \@firstoftwo
 \else \expandafter \@secondoftwo
 \fi
}%
\providecommand \@ifx [1]{%
 \ifx #1\expandafter \@firstoftwo
 \else \expandafter \@secondoftwo
 \fi
}%
\providecommand \natexlab [1]{#1}%
\providecommand \enquote  [1]{``#1''}%
\providecommand \bibnamefont  [1]{#1}%
\providecommand \bibfnamefont [1]{#1}%
\providecommand \citenamefont [1]{#1}%
\providecommand \href@noop [0]{\@secondoftwo}%
\providecommand \href [0]{\begingroup \@sanitize@url \@href}%
\providecommand \@href[1]{\@@startlink{#1}\@@href}%
\providecommand \@@href[1]{\endgroup#1\@@endlink}%
\providecommand \@sanitize@url [0]{\catcode `\\12\catcode `\$12\catcode
  `\&12\catcode `\#12\catcode `\^12\catcode `\_12\catcode `\%12\relax}%
\providecommand \@@startlink[1]{}%
\providecommand \@@endlink[0]{}%
\providecommand \url  [0]{\begingroup\@sanitize@url \@url }%
\providecommand \@url [1]{\endgroup\@href {#1}{\urlprefix }}%
\providecommand \urlprefix  [0]{URL }%
\providecommand \Eprint [0]{\href }%
\providecommand \doibase [0]{https://doi.org/}%
\providecommand \selectlanguage [0]{\@gobble}%
\providecommand \bibinfo  [0]{\@secondoftwo}%
\providecommand \bibfield  [0]{\@secondoftwo}%
\providecommand \translation [1]{[#1]}%
\providecommand \BibitemOpen [0]{}%
\providecommand \bibitemStop [0]{}%
\providecommand \bibitemNoStop [0]{.\EOS\space}%
\providecommand \EOS [0]{\spacefactor3000\relax}%
\providecommand \BibitemShut  [1]{\csname bibitem#1\endcsname}%
\let\auto@bib@innerbib\@empty
\bibitem [{\citenamefont {Anderson}(1958)}]{Anderson1958}%
  \BibitemOpen
  \bibfield  {author} {\bibinfo {author} {\bibfnamefont {P.~W.}\ \bibnamefont
  {Anderson}},\ }\bibfield  {title} {\bibinfo {title} {Absence of diffusion in
  certain random lattices},\ }\href {https://doi.org/10.1103/PhysRev.109.1492}
  {\bibfield  {journal} {\bibinfo  {journal} {Phys. Rev.}\ }\textbf {\bibinfo
  {volume} {109}},\ \bibinfo {pages} {1492} (\bibinfo {year}
  {1958})}\BibitemShut {NoStop}%
\bibitem [{\citenamefont {Hu}\ \emph {et~al.}(2008)\citenamefont {Hu},
  \citenamefont {Strybulevych}, \citenamefont {Page}, \citenamefont
  {Skipetrov},\ and\ \citenamefont {van Tiggelen}}]{Hu2008}%
  \BibitemOpen
  \bibfield  {author} {\bibinfo {author} {\bibfnamefont {H.}~\bibnamefont
  {Hu}}, \bibinfo {author} {\bibfnamefont {A.}~\bibnamefont {Strybulevych}},
  \bibinfo {author} {\bibfnamefont {J.~H.}\ \bibnamefont {Page}}, \bibinfo
  {author} {\bibfnamefont {S.~E.}\ \bibnamefont {Skipetrov}},\ and\ \bibinfo
  {author} {\bibfnamefont {B.~A.}\ \bibnamefont {van Tiggelen}},\ }\bibfield
  {title} {\bibinfo {title} {Localization of ultrasound in a three-dimensional
  elastic network},\ }\href {https://doi.org/10.1038/nphys1101} {\bibfield
  {journal} {\bibinfo  {journal} {Nature Physics}\ }\textbf {\bibinfo {volume}
  {4}},\ \bibinfo {pages} {945} (\bibinfo {year} {2008})}\BibitemShut {NoStop}%
\bibitem [{\citenamefont {Chab\'e}\ \emph {et~al.}(2008)\citenamefont
  {Chab\'e}, \citenamefont {Lemari\'e}, \citenamefont {Gr\'emaud},
  \citenamefont {Delande}, \citenamefont {Szriftgiser},\ and\ \citenamefont
  {Garreau}}]{Chabe2008}%
  \BibitemOpen
  \bibfield  {author} {\bibinfo {author} {\bibfnamefont {J.}~\bibnamefont
  {Chab\'e}}, \bibinfo {author} {\bibfnamefont {G.}~\bibnamefont {Lemari\'e}},
  \bibinfo {author} {\bibfnamefont {B.}~\bibnamefont {Gr\'emaud}}, \bibinfo
  {author} {\bibfnamefont {D.}~\bibnamefont {Delande}}, \bibinfo {author}
  {\bibfnamefont {P.}~\bibnamefont {Szriftgiser}},\ and\ \bibinfo {author}
  {\bibfnamefont {J.~C.}\ \bibnamefont {Garreau}},\ }\bibfield  {title}
  {\bibinfo {title} {Experimental observation of the anderson metal-insulator
  transition with atomic matter waves},\ }\href
  {https://doi.org/10.1103/PhysRevLett.101.255702} {\bibfield  {journal}
  {\bibinfo  {journal} {Phys. Rev. Lett.}\ }\textbf {\bibinfo {volume} {101}},\
  \bibinfo {pages} {255702} (\bibinfo {year} {2008})}\BibitemShut {NoStop}%
\bibitem [{\citenamefont {Ying}\ \emph {et~al.}(2016)\citenamefont {Ying},
  \citenamefont {Gu}, \citenamefont {Chen}, \citenamefont {Wang}, \citenamefont
  {Jin}, \citenamefont {Zhao}, \citenamefont {Zhang},\ and\ \citenamefont
  {Chen}}]{Ying2016}%
  \BibitemOpen
  \bibfield  {author} {\bibinfo {author} {\bibfnamefont {T.}~\bibnamefont
  {Ying}}, \bibinfo {author} {\bibfnamefont {Y.}~\bibnamefont {Gu}}, \bibinfo
  {author} {\bibfnamefont {X.}~\bibnamefont {Chen}}, \bibinfo {author}
  {\bibfnamefont {X.}~\bibnamefont {Wang}}, \bibinfo {author} {\bibfnamefont
  {S.}~\bibnamefont {Jin}}, \bibinfo {author} {\bibfnamefont {L.}~\bibnamefont
  {Zhao}}, \bibinfo {author} {\bibfnamefont {W.}~\bibnamefont {Zhang}},\ and\
  \bibinfo {author} {\bibfnamefont {X.}~\bibnamefont {Chen}},\ }\bibfield
  {title} {\bibinfo {title} {Anderson localization of electrons in single
  crystals: Lixfe7se8},\ }\bibfield  {journal} {\bibinfo  {journal} {Science
  Advances}\ }\textbf {\bibinfo {volume} {2}},\ \href
  {https://doi.org/10.1126/sciadv.1501283} {10.1126/sciadv.1501283} (\bibinfo
  {year} {2016})\BibitemShut {NoStop}%
\bibitem [{\citenamefont {Wiersma}\ \emph {et~al.}(1997)\citenamefont
  {Wiersma}, \citenamefont {Bartolini}, \citenamefont {Lagendijk},\ and\
  \citenamefont {Righini}}]{Wiersma1997}%
  \BibitemOpen
  \bibfield  {author} {\bibinfo {author} {\bibfnamefont {D.~S.}\ \bibnamefont
  {Wiersma}}, \bibinfo {author} {\bibfnamefont {P.}~\bibnamefont {Bartolini}},
  \bibinfo {author} {\bibfnamefont {A.}~\bibnamefont {Lagendijk}},\ and\
  \bibinfo {author} {\bibfnamefont {R.}~\bibnamefont {Righini}},\ }\bibfield
  {title} {\bibinfo {title} {Localization of light in a disordered medium},\
  }\href {https://doi.org/10.1038/37757} {\bibfield  {journal} {\bibinfo
  {journal} {Nature}\ }\textbf {\bibinfo {volume} {390}},\ \bibinfo {pages}
  {671} (\bibinfo {year} {1997})}\BibitemShut {NoStop}%
\bibitem [{\citenamefont {St\"orzer}\ \emph {et~al.}(2006)\citenamefont
  {St\"orzer}, \citenamefont {Gross}, \citenamefont {Aegerter},\ and\
  \citenamefont {Maret}}]{Storzer2006}%
  \BibitemOpen
  \bibfield  {author} {\bibinfo {author} {\bibfnamefont {M.}~\bibnamefont
  {St\"orzer}}, \bibinfo {author} {\bibfnamefont {P.}~\bibnamefont {Gross}},
  \bibinfo {author} {\bibfnamefont {C.~M.}\ \bibnamefont {Aegerter}},\ and\
  \bibinfo {author} {\bibfnamefont {G.}~\bibnamefont {Maret}},\ }\bibfield
  {title} {\bibinfo {title} {Observation of the critical regime near anderson
  localization of light},\ }\href
  {https://doi.org/10.1103/PhysRevLett.96.063904} {\bibfield  {journal}
  {\bibinfo  {journal} {Phys. Rev. Lett.}\ }\textbf {\bibinfo {volume} {96}},\
  \bibinfo {pages} {063904} (\bibinfo {year} {2006})}\BibitemShut {NoStop}%
\bibitem [{\citenamefont {Scheffold}\ \emph {et~al.}(1999)\citenamefont
  {Scheffold}, \citenamefont {Lenke}, \citenamefont {Tweer},\ and\
  \citenamefont {Maret}}]{Scheffold1999}%
  \BibitemOpen
  \bibfield  {author} {\bibinfo {author} {\bibfnamefont {F.}~\bibnamefont
  {Scheffold}}, \bibinfo {author} {\bibfnamefont {R.}~\bibnamefont {Lenke}},
  \bibinfo {author} {\bibfnamefont {R.}~\bibnamefont {Tweer}},\ and\ \bibinfo
  {author} {\bibfnamefont {G.}~\bibnamefont {Maret}},\ }\bibfield  {title}
  {\bibinfo {title} {Localization or classical diffusion of light?},\ }\href
  {https://doi.org/10.1038/18347} {\bibfield  {journal} {\bibinfo  {journal}
  {Nature}\ }\textbf {\bibinfo {volume} {398}},\ \bibinfo {pages} {206}
  (\bibinfo {year} {1999})}\BibitemShut {NoStop}%
\bibitem [{\citenamefont {Scheffold}\ and\ \citenamefont
  {Wiersma}(2013)}]{Scheffold2013}%
  \BibitemOpen
  \bibfield  {author} {\bibinfo {author} {\bibfnamefont {F.}~\bibnamefont
  {Scheffold}}\ and\ \bibinfo {author} {\bibfnamefont {D.}~\bibnamefont
  {Wiersma}},\ }\bibfield  {title} {\bibinfo {title} {Inelastic scattering puts
  in question recent claims of anderson localization of light},\ }\href
  {https://doi.org/10.1038/nphoton.2013.210} {\bibfield  {journal} {\bibinfo
  {journal} {Nature Photonics}\ }\textbf {\bibinfo {volume} {7}},\ \bibinfo
  {pages} {934} (\bibinfo {year} {2013})}\BibitemShut {NoStop}%
\bibitem [{\citenamefont {Sperling}\ \emph {et~al.}(2016)\citenamefont
  {Sperling}, \citenamefont {Schertel}, \citenamefont {Ackermann},
  \citenamefont {Aubry}, \citenamefont {Aegerter},\ and\ \citenamefont
  {Maret}}]{Sperling2016}%
  \BibitemOpen
  \bibfield  {author} {\bibinfo {author} {\bibfnamefont {T.}~\bibnamefont
  {Sperling}}, \bibinfo {author} {\bibfnamefont {L.}~\bibnamefont {Schertel}},
  \bibinfo {author} {\bibfnamefont {M.}~\bibnamefont {Ackermann}}, \bibinfo
  {author} {\bibfnamefont {G.~J.}\ \bibnamefont {Aubry}}, \bibinfo {author}
  {\bibfnamefont {C.~M.}\ \bibnamefont {Aegerter}},\ and\ \bibinfo {author}
  {\bibfnamefont {G.}~\bibnamefont {Maret}},\ }\bibfield  {title} {\bibinfo
  {title} {Can 3d light localization be reached in `white paint'?},\ }\href
  {https://doi.org/10.1088/1367-2630/18/1/013039} {\bibfield  {journal}
  {\bibinfo  {journal} {New Journal of Physics}\ }\textbf {\bibinfo {volume}
  {18}},\ \bibinfo {pages} {013039} (\bibinfo {year} {2016})}\BibitemShut
  {NoStop}%
\bibitem [{\citenamefont {Skipetrov}\ and\ \citenamefont
  {Page}(2016)}]{Skipetrov2016}%
  \BibitemOpen
  \bibfield  {author} {\bibinfo {author} {\bibfnamefont {S.~E.}\ \bibnamefont
  {Skipetrov}}\ and\ \bibinfo {author} {\bibfnamefont {J.~H.}\ \bibnamefont
  {Page}},\ }\bibfield  {title} {\bibinfo {title} {Red light for anderson
  localization},\ }\href {https://doi.org/10.1088/1367-2630/18/2/021001}
  {\bibfield  {journal} {\bibinfo  {journal} {New Journal of Physics}\ }\textbf
  {\bibinfo {volume} {18}},\ \bibinfo {pages} {021001} (\bibinfo {year}
  {2016})}\BibitemShut {NoStop}%
\bibitem [{\citenamefont {Skipetrov}\ and\ \citenamefont
  {Sokolov}(2014)}]{Skipetrov2014}%
  \BibitemOpen
  \bibfield  {author} {\bibinfo {author} {\bibfnamefont {S.~E.}\ \bibnamefont
  {Skipetrov}}\ and\ \bibinfo {author} {\bibfnamefont {I.~M.}\ \bibnamefont
  {Sokolov}},\ }\bibfield  {title} {\bibinfo {title} {Absence of anderson
  localization of light in a random ensemble of point scatterers},\ }\href
  {https://doi.org/10.1103/PhysRevLett.112.023905} {\bibfield  {journal}
  {\bibinfo  {journal} {Phys. Rev. Lett.}\ }\textbf {\bibinfo {volume} {112}},\
  \bibinfo {pages} {023905} (\bibinfo {year} {2014})}\BibitemShut {NoStop}%
\bibitem [{\citenamefont {Rezvani~Naraghi}\ and\ \citenamefont
  {Dogariu}(2016)}]{Naraghi2016}%
  \BibitemOpen
  \bibfield  {author} {\bibinfo {author} {\bibfnamefont {R.}~\bibnamefont
  {Rezvani~Naraghi}}\ and\ \bibinfo {author} {\bibfnamefont {A.}~\bibnamefont
  {Dogariu}},\ }\bibfield  {title} {\bibinfo {title} {Phase transitions in
  diffusion of light},\ }\href {https://doi.org/10.1103/PhysRevLett.117.263901}
  {\bibfield  {journal} {\bibinfo  {journal} {Phys. Rev. Lett.}\ }\textbf
  {\bibinfo {volume} {117}},\ \bibinfo {pages} {263901} (\bibinfo {year}
  {2016})}\BibitemShut {NoStop}%
\bibitem [{\citenamefont {Cobus}\ \emph {et~al.}(2022)\citenamefont {Cobus},
  \citenamefont {Maret},\ and\ \citenamefont {Aubry}}]{Cobus2022}%
  \BibitemOpen
  \bibfield  {author} {\bibinfo {author} {\bibfnamefont {L.~A.}\ \bibnamefont
  {Cobus}}, \bibinfo {author} {\bibfnamefont {G.}~\bibnamefont {Maret}},\ and\
  \bibinfo {author} {\bibfnamefont {A.}~\bibnamefont {Aubry}},\ }\bibfield
  {title} {\bibinfo {title} {Crossover from renormalized to conventional
  diffusion near the three-dimensional anderson localization transition for
  light},\ }\href {https://doi.org/10.1103/PhysRevB.106.014208} {\bibfield
  {journal} {\bibinfo  {journal} {Phys. Rev. B}\ }\textbf {\bibinfo {volume}
  {106}},\ \bibinfo {pages} {014208} (\bibinfo {year} {2022})}\BibitemShut
  {NoStop}%
\bibitem [{\citenamefont {Skipetrov}\ and\ \citenamefont
  {Sokolov}(2015)}]{Skipetrov2015}%
  \BibitemOpen
  \bibfield  {author} {\bibinfo {author} {\bibfnamefont {S.~E.}\ \bibnamefont
  {Skipetrov}}\ and\ \bibinfo {author} {\bibfnamefont {I.~M.}\ \bibnamefont
  {Sokolov}},\ }\bibfield  {title} {\bibinfo {title} {Magnetic-field-driven
  localization of light in a cold-atom gas},\ }\href
  {https://doi.org/10.1103/PhysRevLett.114.053902} {\bibfield  {journal}
  {\bibinfo  {journal} {Phys. Rev. Lett.}\ }\textbf {\bibinfo {volume} {114}},\
  \bibinfo {pages} {053902} (\bibinfo {year} {2015})}\BibitemShut {NoStop}%
\bibitem [{\citenamefont {Celardo}\ \emph {et~al.}(2017)\citenamefont
  {Celardo}, \citenamefont {Angeli},\ and\ \citenamefont
  {Kaiser}}]{Celardo2017}%
  \BibitemOpen
  \bibfield  {author} {\bibinfo {author} {\bibfnamefont {G.~L.}\ \bibnamefont
  {Celardo}}, \bibinfo {author} {\bibfnamefont {M.}~\bibnamefont {Angeli}},\
  and\ \bibinfo {author} {\bibfnamefont {R.}~\bibnamefont {Kaiser}},\
  }\href@noop {} {\bibinfo {title} {Localization of light in subradiant dicke
  states: a mobility edge in the imaginary axis}} (\bibinfo {year} {2017}),\
  \Eprint {https://arxiv.org/abs/arXiv:1702.04506} {arXiv:1702.04506}
  \BibitemShut {NoStop}%
\bibitem [{\citenamefont {Williamson}\ and\ \citenamefont
  {Ruostekoski}(2020)}]{Williamson2020}%
  \BibitemOpen
  \bibfield  {author} {\bibinfo {author} {\bibfnamefont {L.~A.}\ \bibnamefont
  {Williamson}}\ and\ \bibinfo {author} {\bibfnamefont {J.}~\bibnamefont
  {Ruostekoski}},\ }\bibfield  {title} {\bibinfo {title} {Optical response of
  atom chains beyond the limit of low light intensity: The validity of the
  linear classical oscillator model},\ }\href
  {https://doi.org/10.1103/PhysRevResearch.2.023273} {\bibfield  {journal}
  {\bibinfo  {journal} {Phys. Rev. Res.}\ }\textbf {\bibinfo {volume} {2}},\
  \bibinfo {pages} {023273} (\bibinfo {year} {2020})}\BibitemShut {NoStop}%
\bibitem [{\citenamefont {Cipris}\ \emph {et~al.}(2021)\citenamefont {Cipris},
  \citenamefont {Moreira}, \citenamefont {do~Espirito~Santo}, \citenamefont
  {Weiss}, \citenamefont {Villas-Boas}, \citenamefont {Kaiser}, \citenamefont
  {Guerin},\ and\ \citenamefont {Bachelard}}]{Cipris2021}%
  \BibitemOpen
  \bibfield  {author} {\bibinfo {author} {\bibfnamefont {A.}~\bibnamefont
  {Cipris}}, \bibinfo {author} {\bibfnamefont {N.~A.}\ \bibnamefont {Moreira}},
  \bibinfo {author} {\bibfnamefont {T.~S.}\ \bibnamefont {do~Espirito~Santo}},
  \bibinfo {author} {\bibfnamefont {P.}~\bibnamefont {Weiss}}, \bibinfo
  {author} {\bibfnamefont {C.~J.}\ \bibnamefont {Villas-Boas}}, \bibinfo
  {author} {\bibfnamefont {R.}~\bibnamefont {Kaiser}}, \bibinfo {author}
  {\bibfnamefont {W.}~\bibnamefont {Guerin}},\ and\ \bibinfo {author}
  {\bibfnamefont {R.}~\bibnamefont {Bachelard}},\ }\bibfield  {title} {\bibinfo
  {title} {Subradiance with saturated atoms: Population enhancement of the
  long-lived states},\ }\href {https://doi.org/10.1103/PhysRevLett.126.103604}
  {\bibfield  {journal} {\bibinfo  {journal} {Phys. Rev. Lett.}\ }\textbf
  {\bibinfo {volume} {126}},\ \bibinfo {pages} {103604} (\bibinfo {year}
  {2021})}\BibitemShut {NoStop}%
\bibitem [{\citenamefont {Chabanov}\ \emph {et~al.}(2000)\citenamefont
  {Chabanov}, \citenamefont {Stoytchev},\ and\ \citenamefont
  {Genack}}]{Chabanov2000}%
  \BibitemOpen
  \bibfield  {author} {\bibinfo {author} {\bibfnamefont {A.~A.}\ \bibnamefont
  {Chabanov}}, \bibinfo {author} {\bibfnamefont {M.}~\bibnamefont
  {Stoytchev}},\ and\ \bibinfo {author} {\bibfnamefont {A.~Z.}\ \bibnamefont
  {Genack}},\ }\bibfield  {title} {\bibinfo {title} {Statistical signatures of
  photon localization},\ }\href {https://doi.org/10.1038/35009055} {\bibfield
  {journal} {\bibinfo  {journal} {Nature}\ }\textbf {\bibinfo {volume} {404}},\
  \bibinfo {pages} {850} (\bibinfo {year} {2000})}\BibitemShut {NoStop}%
\bibitem [{\citenamefont {Cottier}\ \emph {et~al.}(2019)\citenamefont
  {Cottier}, \citenamefont {Cipris}, \citenamefont {Bachelard},\ and\
  \citenamefont {Kaiser}}]{Cottier2019}%
  \BibitemOpen
  \bibfield  {author} {\bibinfo {author} {\bibfnamefont {F.}~\bibnamefont
  {Cottier}}, \bibinfo {author} {\bibfnamefont {A.}~\bibnamefont {Cipris}},
  \bibinfo {author} {\bibfnamefont {R.}~\bibnamefont {Bachelard}},\ and\
  \bibinfo {author} {\bibfnamefont {R.}~\bibnamefont {Kaiser}},\ }\bibfield
  {title} {\bibinfo {title} {Microscopic and macroscopic signatures of 3d
  anderson localization of light},\ }\href
  {https://doi.org/10.1103/PhysRevLett.123.083401} {\bibfield  {journal}
  {\bibinfo  {journal} {Phys. Rev. Lett.}\ }\textbf {\bibinfo {volume} {123}},\
  \bibinfo {pages} {083401} (\bibinfo {year} {2019})}\BibitemShut {NoStop}%
\bibitem [{\citenamefont {Lehmberg}(1970)}]{Lehmberg1970}%
  \BibitemOpen
  \bibfield  {author} {\bibinfo {author} {\bibfnamefont {R.~H.}\ \bibnamefont
  {Lehmberg}},\ }\bibfield  {title} {\bibinfo {title} {Radiation from an
  $n$-atom system. i. general formalism},\ }\href
  {https://doi.org/10.1103/PhysRevA.2.883} {\bibfield  {journal} {\bibinfo
  {journal} {Phys. Rev. A}\ }\textbf {\bibinfo {volume} {2}},\ \bibinfo {pages}
  {883} (\bibinfo {year} {1970})}\BibitemShut {NoStop}%
\bibitem [{\citenamefont {do~Espirito~Santo}\ \emph {et~al.}(2020)\citenamefont
  {do~Espirito~Santo}, \citenamefont {Weiss}, \citenamefont {Cipris},
  \citenamefont {Kaiser}, \citenamefont {Guerin}, \citenamefont {Bachelard},\
  and\ \citenamefont {Schachenmayer}}]{Santo2020}%
  \BibitemOpen
  \bibfield  {author} {\bibinfo {author} {\bibfnamefont {T.~S.}\ \bibnamefont
  {do~Espirito~Santo}}, \bibinfo {author} {\bibfnamefont {P.}~\bibnamefont
  {Weiss}}, \bibinfo {author} {\bibfnamefont {A.}~\bibnamefont {Cipris}},
  \bibinfo {author} {\bibfnamefont {R.}~\bibnamefont {Kaiser}}, \bibinfo
  {author} {\bibfnamefont {W.}~\bibnamefont {Guerin}}, \bibinfo {author}
  {\bibfnamefont {R.}~\bibnamefont {Bachelard}},\ and\ \bibinfo {author}
  {\bibfnamefont {J.}~\bibnamefont {Schachenmayer}},\ }\bibfield  {title}
  {\bibinfo {title} {Collective excitation dynamics of a cold atom cloud},\
  }\href {https://doi.org/10.1103/PhysRevA.101.013617} {\bibfield  {journal}
  {\bibinfo  {journal} {Phys. Rev. A}\ }\textbf {\bibinfo {volume} {101}},\
  \bibinfo {pages} {013617} (\bibinfo {year} {2020})}\BibitemShut {NoStop}%
\bibitem [{\citenamefont {Lassègues}\ \emph {et~al.}(2022)\citenamefont
  {Lassègues}, \citenamefont {Biscassi}, \citenamefont {Morisse},
  \citenamefont {Cidrim}, \citenamefont {Dias}, \citenamefont {Eneriz},
  \citenamefont {Teixeira}, \citenamefont {Kaiser}, \citenamefont {Bachelard},\
  and\ \citenamefont {Hugbart}}]{Lassegues2023}%
  \BibitemOpen
  \bibfield  {author} {\bibinfo {author} {\bibfnamefont {P.}~\bibnamefont
  {Lassègues}}, \bibinfo {author} {\bibfnamefont {M.~A.~F.}\ \bibnamefont
  {Biscassi}}, \bibinfo {author} {\bibfnamefont {M.}~\bibnamefont {Morisse}},
  \bibinfo {author} {\bibfnamefont {A.}~\bibnamefont {Cidrim}}, \bibinfo
  {author} {\bibfnamefont {P.~G.~S.}\ \bibnamefont {Dias}}, \bibinfo {author}
  {\bibfnamefont {H.}~\bibnamefont {Eneriz}}, \bibinfo {author} {\bibfnamefont
  {R.~C.}\ \bibnamefont {Teixeira}}, \bibinfo {author} {\bibfnamefont
  {R.}~\bibnamefont {Kaiser}}, \bibinfo {author} {\bibfnamefont
  {R.}~\bibnamefont {Bachelard}},\ and\ \bibinfo {author} {\bibfnamefont
  {M.}~\bibnamefont {Hugbart}},\ }\href@noop {} {\bibinfo {title} {From
  classical to quantum loss of light coherence}} (\bibinfo {year} {2022}),\
  \Eprint {https://arxiv.org/abs/arXiv:2210.01003} {arXiv:2210.01003}
  \BibitemShut {NoStop}%
\bibitem [{\citenamefont {Pucci}\ \emph {et~al.}(2017)\citenamefont {Pucci},
  \citenamefont {Roy}, \citenamefont {do~Espirito~Santo}, \citenamefont
  {Kaiser}, \citenamefont {Kastner},\ and\ \citenamefont
  {Bachelard}}]{Pucci2017}%
  \BibitemOpen
  \bibfield  {author} {\bibinfo {author} {\bibfnamefont {L.}~\bibnamefont
  {Pucci}}, \bibinfo {author} {\bibfnamefont {A.}~\bibnamefont {Roy}}, \bibinfo
  {author} {\bibfnamefont {T.~S.}\ \bibnamefont {do~Espirito~Santo}}, \bibinfo
  {author} {\bibfnamefont {R.}~\bibnamefont {Kaiser}}, \bibinfo {author}
  {\bibfnamefont {M.}~\bibnamefont {Kastner}},\ and\ \bibinfo {author}
  {\bibfnamefont {R.}~\bibnamefont {Bachelard}},\ }\bibfield  {title} {\bibinfo
  {title} {Quantum effects in the cooperative scattering of light by atomic
  clouds},\ }\href {https://doi.org/10.1103/PhysRevA.95.053625} {\bibfield
  {journal} {\bibinfo  {journal} {Phys. Rev. A}\ }\textbf {\bibinfo {volume}
  {95}},\ \bibinfo {pages} {053625} (\bibinfo {year} {2017})}\BibitemShut
  {NoStop}%
\bibitem [{\citenamefont {Plankensteiner}\ \emph {et~al.}(2022)\citenamefont
  {Plankensteiner}, \citenamefont {Hotter},\ and\ \citenamefont
  {Ritsch}}]{Plankensteiner2022}%
  \BibitemOpen
  \bibfield  {author} {\bibinfo {author} {\bibfnamefont {D.}~\bibnamefont
  {Plankensteiner}}, \bibinfo {author} {\bibfnamefont {C.}~\bibnamefont
  {Hotter}},\ and\ \bibinfo {author} {\bibfnamefont {H.}~\bibnamefont
  {Ritsch}},\ }\bibfield  {title} {\bibinfo {title} {{QuantumCumulants}.jl: A
  julia framework for generalized mean-field equations in open quantum
  systems},\ }\href {https://doi.org/10.22331/q-2022-01-04-617} {\bibfield
  {journal} {\bibinfo  {journal} {Quantum}\ }\textbf {\bibinfo {volume} {6}},\
  \bibinfo {pages} {617} (\bibinfo {year} {2022})}\BibitemShut {NoStop}%
\bibitem [{\citenamefont {Guerin}\ and\ \citenamefont
  {Kaiser}(2017)}]{Guerin2017}%
  \BibitemOpen
  \bibfield  {author} {\bibinfo {author} {\bibfnamefont {W.}~\bibnamefont
  {Guerin}}\ and\ \bibinfo {author} {\bibfnamefont {R.}~\bibnamefont
  {Kaiser}},\ }\bibfield  {title} {\bibinfo {title} {Population of collective
  modes in light scattering by many atoms},\ }\href
  {https://doi.org/10.1103/PhysRevA.95.053865} {\bibfield  {journal} {\bibinfo
  {journal} {Phys. Rev. A}\ }\textbf {\bibinfo {volume} {95}},\ \bibinfo
  {pages} {053865} (\bibinfo {year} {2017})}\BibitemShut {NoStop}%
\bibitem [{\citenamefont {Moreira}\ \emph {et~al.}(2019)\citenamefont
  {Moreira}, \citenamefont {Kaiser},\ and\ \citenamefont
  {Bachelard}}]{Moreira2019}%
  \BibitemOpen
  \bibfield  {author} {\bibinfo {author} {\bibfnamefont {N.~A.}\ \bibnamefont
  {Moreira}}, \bibinfo {author} {\bibfnamefont {R.}~\bibnamefont {Kaiser}},\
  and\ \bibinfo {author} {\bibfnamefont {R.}~\bibnamefont {Bachelard}},\
  }\bibfield  {title} {\bibinfo {title} {Localization versus subradiance in
  three-dimensional scattering of light},\ }\href
  {https://doi.org/10.1209/0295-5075/127/54003} {\bibfield  {journal} {\bibinfo
   {journal} {{EPL} (Europhysics Letters)}\ }\textbf {\bibinfo {volume}
  {127}},\ \bibinfo {pages} {54003} (\bibinfo {year} {2019})}\BibitemShut
  {NoStop}%
\bibitem [{\citenamefont {Rubies-Bigorda}\ \emph {et~al.}(2023)\citenamefont
  {Rubies-Bigorda}, \citenamefont {Ostermann},\ and\ \citenamefont
  {Yelin}}]{Rubies2023}%
  \BibitemOpen
  \bibfield  {author} {\bibinfo {author} {\bibfnamefont {O.}~\bibnamefont
  {Rubies-Bigorda}}, \bibinfo {author} {\bibfnamefont {S.}~\bibnamefont
  {Ostermann}},\ and\ \bibinfo {author} {\bibfnamefont {S.~F.}\ \bibnamefont
  {Yelin}},\ }\bibfield  {title} {\bibinfo {title} {Dynamic population of
  multiexcitation subradiant states in incoherently excited atomic arrays},\
  }\href {https://doi.org/10.1103/PhysRevA.107.L051701} {\bibfield  {journal}
  {\bibinfo  {journal} {Phys. Rev. A}\ }\textbf {\bibinfo {volume} {107}},\
  \bibinfo {pages} {L051701} (\bibinfo {year} {2023})}\BibitemShut {NoStop}%
\bibitem [{\citenamefont {Fayard}\ \emph {et~al.}(2021)\citenamefont {Fayard},
  \citenamefont {Henriet}, \citenamefont {Asenjo-Garcia},\ and\ \citenamefont
  {Chang}}]{Fayard2021}%
  \BibitemOpen
  \bibfield  {author} {\bibinfo {author} {\bibfnamefont {N.}~\bibnamefont
  {Fayard}}, \bibinfo {author} {\bibfnamefont {L.}~\bibnamefont {Henriet}},
  \bibinfo {author} {\bibfnamefont {A.}~\bibnamefont {Asenjo-Garcia}},\ and\
  \bibinfo {author} {\bibfnamefont {D.~E.}\ \bibnamefont {Chang}},\ }\bibfield
  {title} {\bibinfo {title} {Many-body localization in waveguide quantum
  electrodynamics},\ }\href {https://doi.org/10.1103/PhysRevResearch.3.033233}
  {\bibfield  {journal} {\bibinfo  {journal} {Phys. Rev. Res.}\ }\textbf
  {\bibinfo {volume} {3}},\ \bibinfo {pages} {033233} (\bibinfo {year}
  {2021})}\BibitemShut {NoStop}%
\end{thebibliography}%

\end{document}